\documentclass[aps,prb,twocolumn,showpacs,psfig,superscriptaddress,longbibliography]{revtex4-1}
\usepackage{amsfonts}
\usepackage{mathrsfs}
\usepackage{amsmath}
\usepackage{color}
\usepackage{natbib}
\usepackage{textcomp}
\usepackage{graphicx}
\usepackage{bm}
\usepackage{amssymb}

\usepackage{xspace}
\usepackage{epstopdf}
\usepackage{dcolumn}
\usepackage{longtable}
\usepackage{multirow}
\usepackage[colorlinks=true, letterpaper=true, pdfstartview=FitV, linkcolor=blue, citecolor=blue, urlcolor=blue]{hyperref}
\usepackage{float}

\makeatletter

\newcommand{\Rmnum}[1]{\expandafter\@slowromancap\romannumeral #1@}
\makeatother

\begin{document}

\title{Two-Dimensional Magnetic Semiconductors with Room Curie Temperatures}
 \author{Jing-Yang You}
 \affiliation{School of Physical Sciences, University of Chinese Academy of Sciences, Beijing 100049, China}

 \author{Zhen Zhang}
 \affiliation{School of Physical Sciences, University of Chinese Academy of Sciences, Beijing 100049, China}

 \author{Xue-Juan Dong}
 \affiliation{School of Physical Sciences, University of Chinese Academy of Sciences, Beijing 100049, China}

 \author{Bo Gu}
 \email{gubo@ucas.ac.cn}
 \affiliation{Kavli Institute for Theoretical Sciences, and CAS Center for Excellence in Topological Quantum Computation, University of Chinese Academy of Sciences, Beijng 100190, China}
\affiliation{Physical Science Laboratory, Huairou National Comprehensive Science Center, 101400 Beijing, China}

 \author{Gang Su}
 \email{gsu@ucas.ac.cn}
 \affiliation{School of Physical Sciences, University of Chinese Academy of Sciences, Beijing 100049, China}
 \affiliation{Kavli Institute for Theoretical Sciences, and CAS Center for Excellence in Topological Quantum Computation, University of Chinese Academy of Sciences, Beijng 100190, China}
\affiliation{Physical Science Laboratory, Huairou National Comprehensive Science Center, 101400 Beijing, China}

\begin{abstract}
We propose two-dimensional (2D) Ising-type ferromagnetic semiconductors TcSiTe$_3$, TcGeSe$_3$ and TcGeTe$_3$ with high Curie temperatures around 200 $\sim$ 500 K. Owing to large spin-orbit couplings, the large magnetocrystalline anisotropy energy (MAE), large anomalous Hall conductivity, and large magneto-optical Kerr effect were discovered in these intriguing 2D materials. By comparing all possible 2D MGeTe$_3$ materials (M = 3d, 4d, 5d transition metals),  we found a large orbital moment around 0.5 $\mu_B$ per atom and a large MAE for TcGeTe$_3$. The large orbital moments are unveiled to be from the comparable crystal fields and electron correlations in these Tc-based 2D materials. The microscopic mechanism of the high Curie temperature is also addressed. Our findings expose the unique magnetic behaviors of 2D Tc-based materials, and present a new family of 2D ferromagnetic semiconductors with large MAE and Kerr rotation angles that would have wide applications in designing spintronic devices.
\end{abstract}
\maketitle


\section{Introduction}

Spin-orbit coupling (SOC) describes the relativistic interaction between the spin and orbital momentum of electrons~\cite{Soumyanarayanan2016}. SOC can drive rich phenomena, such as magnetic anisotropy~\cite{Johnson1996}, spin relaxation~\cite{Wu2010}, magnetic damping~\cite{Millsa}, anisotropic magnetoresistance~\cite{McGuire1976}, and anomalous Hall effect~\cite{Nagaosa2010}. Recently, a term spin-orbitronics is proposed to cover the expanding research filed, where SOC is a key concept~\cite{Manchon2015,Soumyanarayanan2016,Manchon2017}. Combining strong SOC and magnetism, many intriguing physical phenomena can be achieved, including current-driven magnetization reversal~\cite{Miron2011,Liu2012,Garello2014}, domain wall propagation~\cite{Miron2011a,Yang2015}, current-driven skyrmion motion~\cite{Jiang2015,Woo2016,Jiang2016}, etc. Transition metals are usually candidates to realize these phenomena and play important roles in spin-orbitronics.

Magnetic anisotropy is one of the fundamental properties of magnetic materials. It is a key issue in recent advances in two-dimensional (2D) magnetic semiconductors~\cite{Burch2018,Huang2017,Gong2017,Dong2019}. According to Mermin-Wagner theorem~\cite{Mermin1966}, at finite temperatures, the quantum spin-S Heisenberg model with isotropic and finite-range exchange interactions in 1D or 2D lattices can be neither ferro- nor anti-ferromagnetism. Thus, for stabilizing long-range ferromagnetic order in 2D magnetic semiconductors at finite temperature, a large magnetic anisotropy, which makes the systems away from the isotropic Heisenberg model, is extremely important.

Magneto-optical Kerr effect (MOKE), that is closely related to SOC, is a basic magneto-optic effect. It describes that the plane-polarized light reflected from a magnetized material becomes elliptically polarized, and the plane of polarization is rotated. MOKE is widely used to probe the electronic structure of magnetic materials. Many exciting phenomena related to MOKE have been discovered, such as quantum
confinement effects~\cite{Suzuki1992}, oscillations of the Kerr rotation with magnetic layer thickness~\cite{Bennett1990} and strong correlations between MOKE and magnetic anisotropies~\cite{Weller1993}. Due to the application of MOKE to the readout process in magneto-optical (MO) storage devices, many efforts have been devoted to searching for materials with large Kerr rotation angles.

In this paper, we propose three stable 2D ferromagnetic semiconductors TcSiTe$_3$, TcGeSe$_3$ and TcGeTe$_3$, which share the same crystal structure as the recently discovered 2D magnetic semiconductor CrGeTe$_3$~\cite{Gong2017}. These Tc-based 2D materials have not been observed experimentally yet. The Monte Carlo simulations give Curie temperatures 538 K, 212 K and 187 K for TcSiTe$_3$, TcGeSe$_3$ and TcGeTe$_3$ monolayers, respectively, which are much higher than the Curie temperature in CrGeTe$_3$. The calculations show that these Tc-based materials have spin moment about 2 $\mu_B$ and an extraordinarily large orbital moment about 0.5 $\mu_B$ per Tc atom. The large orbital moment comes from the partially occupied $d$ orbitals, and the partial occupation of $d$ orbitals is due to the comparable crystal fields and electron correlations in these Tc-based 2D materials. As a result, a large SOC is obtained in these materials. Due to the large SOC, a large magnetocrystalline anisotropy energy (MAE) is formed, indicating the Ising behavior of these 2D materials with out-of-plane magnetization. In addition, a large Kerr rotation angle about 3.6 degree is achieved in these Tc-based materials, which is much larger than the value of 0.8 degree in metal Fe. Large anomalous Hall conductivity of about 7.5$\times$10$^2$ ($\Omega$$\cdot$cm)$^{-1}$ in p-type TcGeTe$_3$ and 1.1$\times$10$^3$ ($\Omega$$\cdot$cm)$^{-1}$ in n-type TcGeTe$_3$ is comparable to the anomalous Hall conductivity of 7.5$\times$10$^2$ ($\Omega$$\cdot$cm)$^{-1}$ in bulk Fe~\cite{Yao2004,Wang2006} and 4.8$\times$10$^2$ ($\Omega$$\cdot$cm)$^{-1}$ in bulk Ni~\cite{Wang2007}.  The microscopic
mechanism of high Curie temperature in these Tc-based materials is also discussed.

\section{Computational method}
Our first-principles calculations were based on the density-functional theory (DFT) as implemented in the Vienna \textit{ab initio} simulation package (VASP)~\cite{Kresse1996}, using the projector augmented wave method~\cite{Bloechl1994}. The generalized gradient approximation (GGA) with Perdew-Burke-Ernzerhof~\cite{Perdew1996} realization was adopted for the exchange-correlation functional. We take the on-site Hubbard interaction $U$ = 2.3 eV and Hund coupling $J$ = 0.3 eV~\cite{Mravlje2012} for considering electron correlation of 4$d$ electrons of Tc atoms, and the effective $U_{eff}=U-J=2$ eV, because the reasonable $U_{eff}$ is about 2 eV for 4$d$ electrons. The plane-wave cutoff energy was set to 550 eV. The Monkhorst-Pack $k$-point mesh~\cite{Monkhorst1976} of size $13\times13\times 1$ was used for the BZ sampling. The structure relaxation considering both the atomic positions and lattice vectors was performed by the conjugate gradient (CG) scheme until the maximum force on each atom was less than 0.0001 eV/{\AA}, and the total energy was converged to $10^{-8}$ eV with Gaussian smearing method. To avoid unnecessary interactions between the monolayer and its periodic images, the vacuum layer is set to 15 {\AA}.  The phonon frequencies were calculated using a finite displacement approach as implemented in the PHONOPY code~\cite{Togo2015}, in which a 3$\times$3$\times$1 supercell and a displacement of 0.01 {\AA} from the equilibrium atomic positions are employed. Wannier90 code~\cite{Mostofi2014} is used to construct an effective tight-binding Hamiltonian and to calculate the optic conductivity and the anomalous Hall conductivity.

\begin{figure}[tbhp]
  \centering
  \includegraphics[scale=0.48,angle=0]{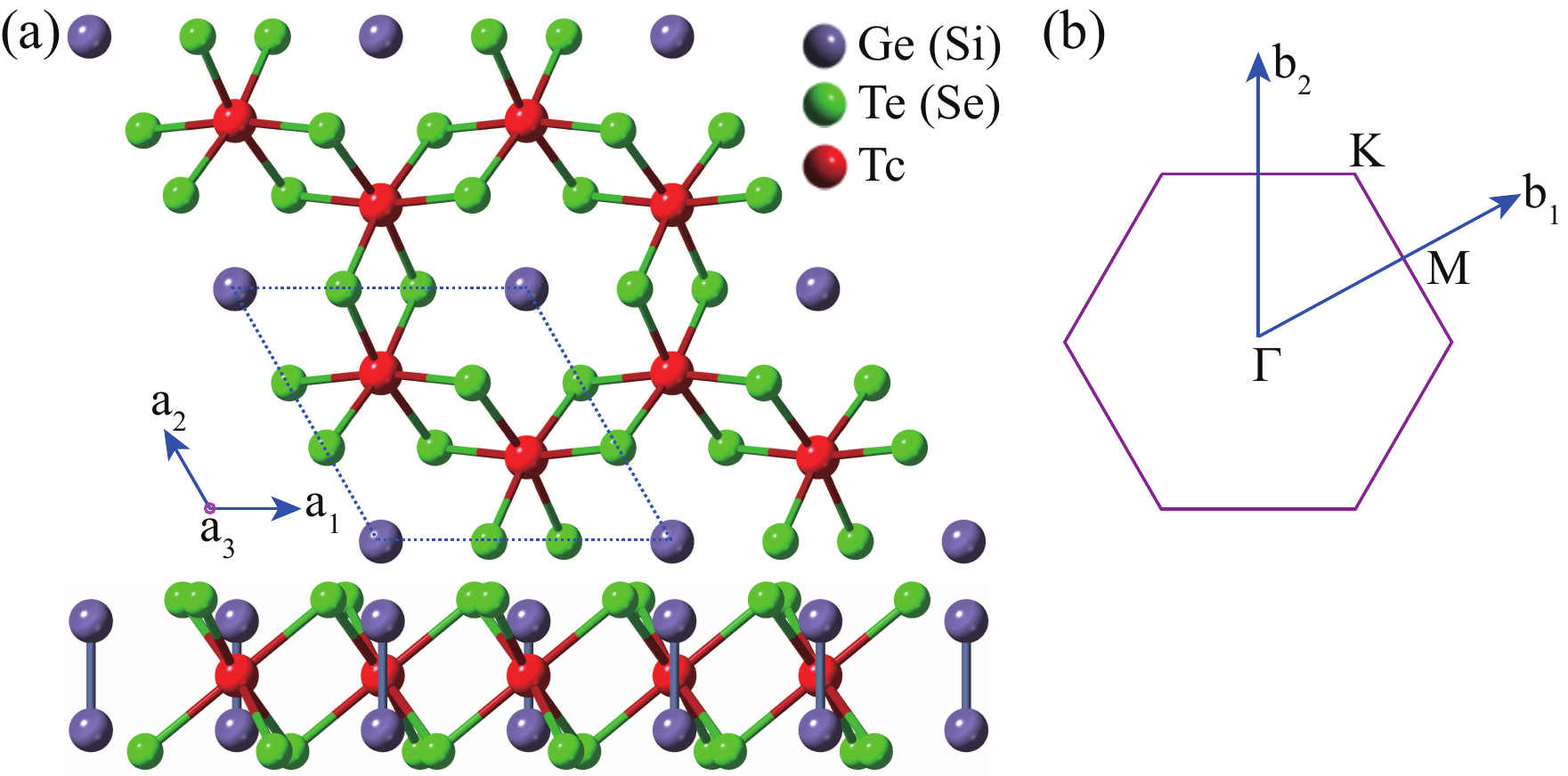}\\
  \caption{(a) Top and side views of the crystal structure of TcSiTe$_3$, TcGeSe$_3$ and TcGeTe$_3$ monolayers. The primitive cell is noted by a dashed line box. Tc atoms form a honeycomb lattice. (b) The first Brillouin zone.}\label{fig1}
\end{figure}

\section{Results}
The crystal structure of TcSiTe$_3$, TcGeSe$_3$ and TcGeTe$_3$ monolayers from the prototype CrGeTe$_3$ monolayer is depicted in Fig.~\ref{fig1}(a), where the spcae group is $P\bar{3}1m$ (No.191). To determine the ground state of TcSiTe$_3$, TcGeSe$_3$ and TcGeTe$_3$ monolayers, in the absence of SOC, we calculated the total energy for ferromagnetic (FM) and antiferromagnetic (AFM) configurations as a function of lattice constant, and found that the FM state has an energy lower than AFM state. The optimized lattice constants of 2D TcSiTe$_3$, TcGeSe$_3$ and TcGeTe$_3$ are calculated as 6.821 {\AA}, 6.379 {\AA} and 7.029 {\AA}, respectively, which are reasonable according to the radius of atoms.

To confirm the stability of these three monolayers, their phonon spectra have been calculated. There is no imaginary frequency mode in the whole Brillouin zone as shown in Figs. S1(a), (b) and (c) (Supplemental Material), indicating that they are kinetically stable. We have checked the stability of these structures with different $U_{eff}$ (1 eV and 3 eV), and the results show that these structures are always stable. To further examine the thermal stability, we performed ab initio molecular dynamics simulations using a $4\times4\times1$ supercell containing 160 atoms. After being heated at 300 K and 500 K for 6 ps with a time step of 3 fs, only little structural and energetic changes occur as shown in Figs. S2(d), (e) and (f) (Supplemental Material), implying that TcSiTe$_3$, TcGeSe$_3$ and TcGeTe$_3$ monolayers are dynamically stable.

\begin{figure}[tbhp]
  \centering
  \includegraphics[scale=0.8,angle=0]{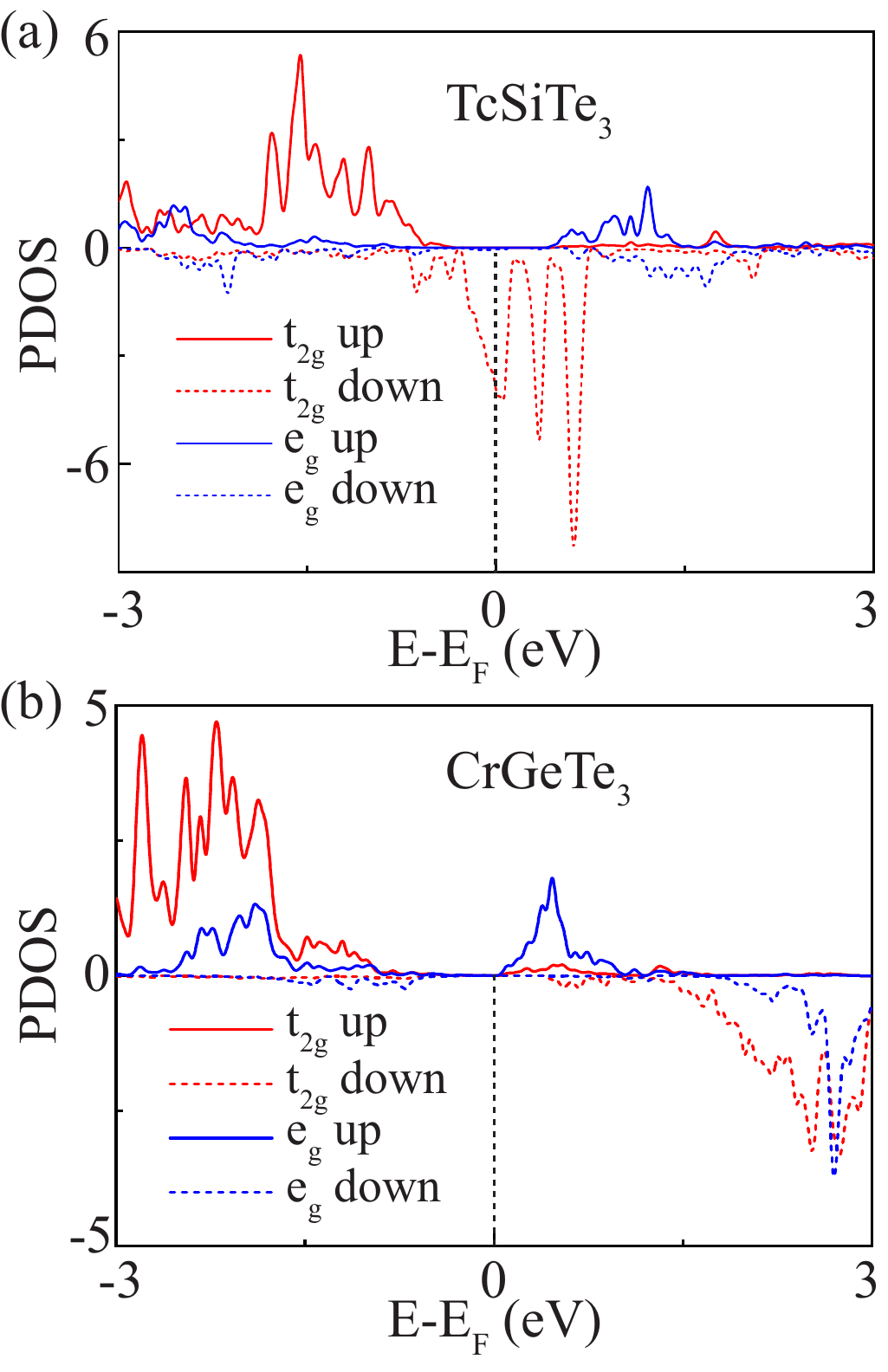}\\
  \caption{The partial density of states (PDOS)  of (a) TcSiTe$_3$ and (b) CrGeTe$_3$ monolayers, calculated by GGA+U method.}\label{fig2}
\end{figure}

\begin{figure*}[tbhp]
  \centering
  \includegraphics[scale=0.85,angle=0]{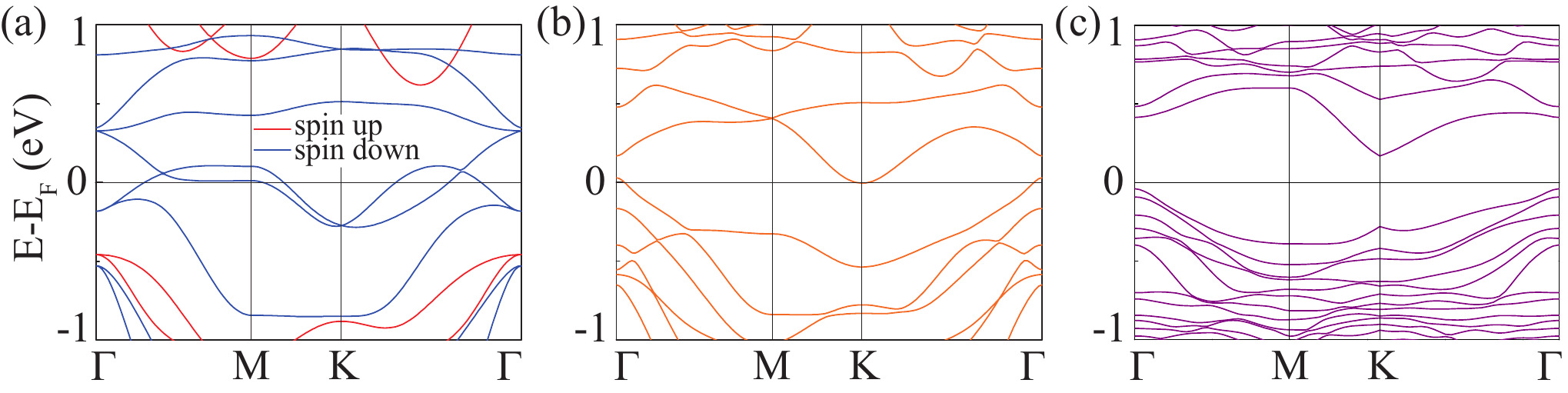}\\
  \caption{The electronic band structures of TcSiTe$_3$ calculated by (a) GGA+U, (b) GGA+SOC+U, and (c) HSE06 methods.}\label{fig3}
\end{figure*}

\begin{table*}[t]
	\caption{The total energy $E_\mathrm{tot}$ per unit cell for TcSiTe$_3$, TcGeSe$_3$ and TcGeTe$_3$ monolayers (in meV, relative to $E_\mathrm{tot}$ of FM$^z$ ground state) for several spin configurations of Tc atoms (see Fig. 4), calculated by GGA+SOC+U method. The spin moment $\langle S\rangle$ and orbital moment $\langle O\rangle$ (in unit of $\mu_B$), single ion anisotropy (SIA, in meV) between the out-of-plane and in-plane FM configurations, exchange interaction $J$ (in meV), and Curie temperature $T_{Curie}$ (in K) are calculated. The CrGeTe$_3$ monolayer is also calculated with the experimental lattice constant~\cite{Gong2017} for comparison.}\label{tab:magnet}
	\begin{tabular}{ccccccccccccc}
		\hline
		                     & FM$^z$ &FM$^x$ &FM$^y$ & NAFM$^z$ & SAFM$^z$ & ZAFM$^z$   &PM  &$\langle S\rangle$  &$\langle O\rangle$  &SIA  &$J$ &$T_{Curie}(K)$\\
		\hline \hline
		TcSiTe$_3$        &0.0     &52.1     &25.9    &310.4     &165.1     &122.0        &2000  &1.866 &0.540 &-42.5  &7.625 &538\\
           TcGeSe$_3$      &0.0      &53.6     &211.9  &295.3     &95.9      &138.0      &2065  &1.884 &0.515 &-37.7  &2.997 &212\\
		TcGeTe$_3$      &0.0    &112.1    &289.7  &277.0     &84.7      &114.2      &2300  &1.991  &0.562 &-26.5  &2.647 &187\\
           CrGeTe$_3$      &0.0     &3.7    &3.8        &143.6     &84.7      &4.3       &6000  &3.614 &0.004 &0.032  &0.066 &19\\
        \hline	
	\end{tabular}
\end{table*}

The structural stabilities of TcSiTe$_3$, TcGeSe$_3$ and TcGeTe$_3$ are also examined by the formation energy, which is calculated by $E_f = E(TcAB_3) - E(Tc) - E(A) - 3E(B)$, where $E(Tc)$, $E(A)$ and $E(B)$ are the total energies of the bulk Tc, Ge (Si), and Se (Te) crystals, respectively. The obtained negative values $E_f$= -0.856 eV, -1.582 eV and -0.694 eV for TcSiTe$_3$, TcGeSe$_3$ and TcGeTe$_3$, respectively. For CrGeTe$_3$ monolayer, which was discovered in recent experiment~\cite{Gong2017}, the formation energy was calculated as  -1.140 eV by the same method. The comparable formation energy of TcSiTe$_3$, TcGeSe$_3$ and TcGeTe$_3$ with CrGeTe$_3$, suggests that these Tc-based materials may also be feasible in experiment.

Since TcSiTe$_3$, TcGeSe$_3$ and TcGeTe$_3$ have  similar properties, we will take TcSiTe$_3$ in the following analysis, and the calculated results of TcGeSe$_3$ and TcGeTe$_3$ are shown in supplemental material. The partial density of states (PDOS) of TcSiTe$_3$ monolayer was calculated by GGA+U method, as illustrated in Fig.~\ref{fig2}(a). Because of the octahedral crystal field for Tc atom, the d orbitals of the Tc atoms are split into threefold $t_{2g}$ orbitals and twofold $e_g$ orbitals. For Tc$^{2+}$ (4d$^5$) in the TcSiTe$_3$ monolayer, the spin moment S = 2 $\mu_B$ and orbital moment L = 0.6 $\mu_B$ are obtained. The results can be understood by the following electron configurations: 3 spin up electrons and 0.6 spin down electrons occupy t$_{2g}$ orbitals, and 0.5 spin up electrons and 0.9 spin down electrons occupy $e_g$ orbitals. It reflects that the crystal field and Coulomb interaction U are comparable for 4d electrons of Tc. In contrast, for Cr$^{2+}$ (3d$^4$) in CrGeTe$_3$ monolayer (U is taken as 4 eV), 3 spin up electrons are in $t_{2g}$ orbitals and 1 spin up electron is in $e_g$ orbitals, which gives rise to the spin moment S = 4 $\mu_B$ and orbital moment L = 0, as shown in Fig.~\ref{fig2}(b). It means that the Coulomb interaction U is much larger than the crystal field for 3d electrons of Cr. These results can also be obtained by integrating the total density of states below the Fermi level for spin up and spin down electrons for $t_{2g}$ and $e_g$ orbitals, respectively. The SOC is calculated by $H_\text{SOC}=\lambda \textbf{S} \cdot \textbf{L}$, where $\lambda$ is the coefficient of SOC, $\textbf{S}$ and $\textbf{L}$ represent the spin and orbital moment operators, respectively.  Because of the large $\lambda$, which is related to the atomic number and large orbital moment L, a much larger SOC is expected in TcSiTe$_3$ monolayer than that in CrGeTe$_3$ monolayer.

\begin{figure}[tbhp]
  \centering
  \includegraphics[scale=0.42,angle=0]{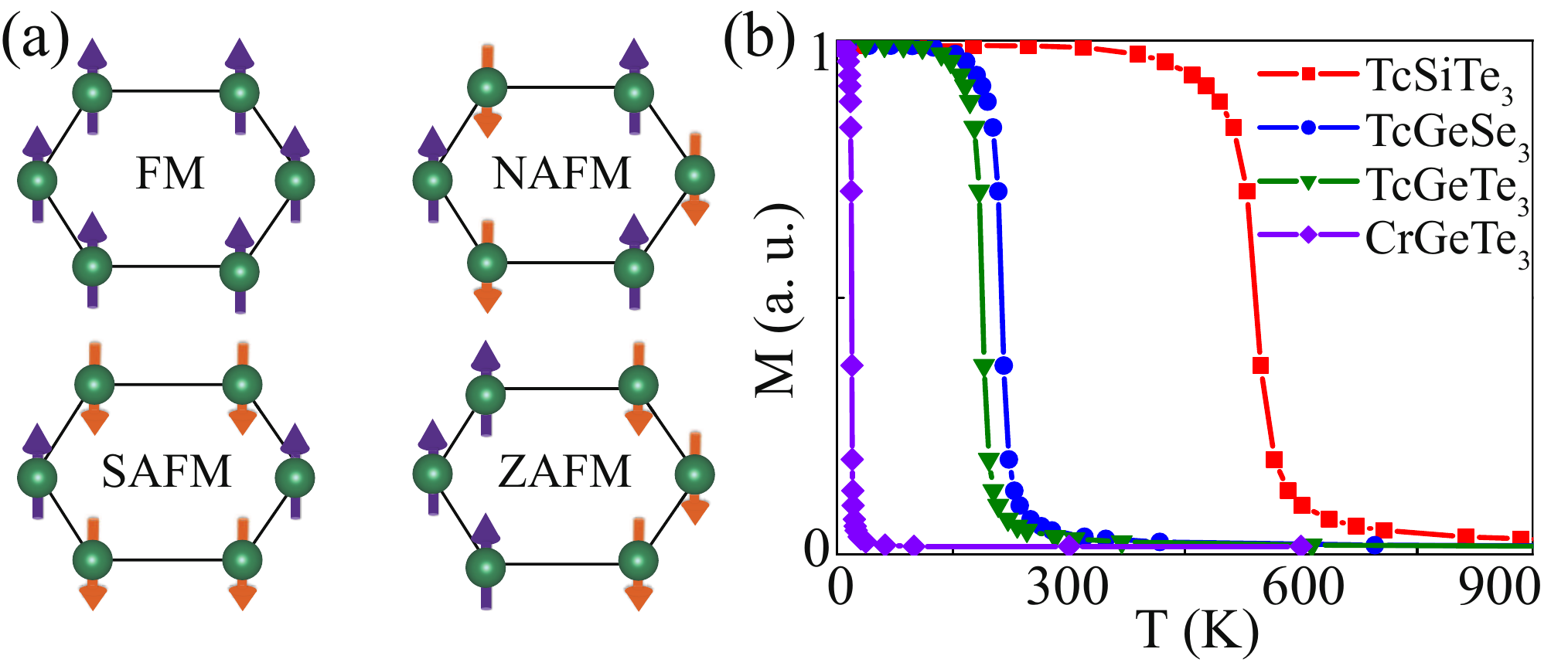}\\
  \caption{(a) Possible spin configurations of Tc atoms on honeycomb lattice: FM, N$\acute{e}$el AFM (NAFM), stripe AFM (SAFM), and zigzag AFM (ZAFM). (b) Temperature dependence of the normalized magnetic moment of TcSiTe$_3$, TcGeSe$_3$, TcGeTe$_3$ and CrGeTe$_3$ monolayers by Monte Carlo simulations.}\label{fig4}
\end{figure}

The electronic band structure of TcSiTe$_3$ monolayer was calculated by GGA+U method, as shown in Fig.~\ref{fig3}(a). It is a Weyl half-metal, where only one species of electron spin appears at Fermi level. At the high-symmetry lines $\Gamma-K$ and $\Gamma-M$, there exist Weyl nodes. Considering the inversion and $C_{3v}$ symmetries, there are total 12 Weyl points within the BZ.
To demonstrate the effect of SOC, the electronic band structure was calculated by GGA+SOC+U method, as plotted in Fig.~\ref{fig3}(b). As a result of including SOC, the band gap was opened for TcSiTe$_3$ monolayer. To correctly estimate the band gap, since the GGA-type calculations usually underestimate the band gap, the hybrid functional method HSE06 was also employed. The HSE06 calculation shows that the band gap of TcSiTe$_3$ monolayer becomes 0.4 eV, as shown in Fig.~\ref{fig3}(c). The details of electronic band structures were given in Supplemental Material.

Due to the large SOC, a large magnetic anisotropy is highly expected for TcSiTe$_3$, TcGeSe$_3$ and TcGeTe$_3$ monolayers. To study magnetic anisotropy in these monolayers, we calculated the total energy with possible spin configurations of Tc atoms on honeycomb lattice, including paramagnetic (PM), ferromagnetic (FM), N$\acute{e}$el antiferromagnetic (NAFM), stripe AFM (SAFM), and zigzag AFM (ZAFM) configurations, as shown in Fig.~\ref{fig4}(a). CrGeTe$_3$ monolayer was also calculated in the same way for comparison. The results are summarized in Table~\ref{tab:magnet}. One can observe that the out-of-plane FM (FM$^z$) state has the lowest energy among the possible spin configurations. The magnetic anisotropy between the in-plane magnetic configuration FM$^x$ and FM$^y$ and the out-of-plane magnetic configurations FM$^z$ in Tc-base materials is extraordinarily lager than that in CrGeTe$_3$, as noted in Table~\ref{tab:magnet}. We further calculated the energies for FM configurations by rotating the magnetic direction deviated from the z-axis, and found that the FM$^z$ state is the most energetically favorable, which shows an Ising behavior of TcSiTe$_3$, TcGeSe$_3$ and TcGeTe$_3$ monolayer. The calculation reveals that the magnetic anisotropy originates from the single-ion anisotropy (SIA), the latter can be calculated by four ordered spin states~\cite{Xiang2013}. The results in Table \ref{tab:magnet} show that, for TcSiTe$_3$, TcGeSe$_3$ and TcGeTe$_3$ monolayers, SIA are found to be negative and large, which determines a strong Ising-type behavior with out-of-plane magnetization. For CrGeTe$_3$ monolayer, SIA is negligible and approaches to zero, which means the Heisenberg-like behavior with weak magnetic anisotropy.

Thus, the magnetism in TcSiTe$_3$, TcGeSe$_3$ and TcGeTe$_3$ monolayers can be described by the Ising-type Hamiltonian $H_{spin}=-\sum_{\langle i,j\rangle}JS_{i}^{z}S_{j}^{z}$,  where $J$ represents the nearest-neighbor exchange integral, $S_{i,j}^{z}$ is the z-component of spin operator, and ${\langle i,j\rangle}$ denotes the summation over the nearest neighbors. $J$ can be determined by the difference of energies between FM$^z$ configuration and AFM configuration, which possesses the lowest energy among those AFM configurations. In our cases, the ZAFM$^z$ configuration has the lowest energy for TcSiTe$_3$ and CrGeTe$_3$ monolayers, and the SAFM$^z$ configuration owns the lowest energy for TcGeSe$_3$ and TcGeTe$_3$ monolayers, as shown in Table \ref{tab:magnet}. As a result, $J$ was estimated to be 7.625 meV, 2.997 meV, 2.647 meV  and 0.066 meV for TcSiTe$_3$, TcGeSe$_3$, TcGeTe$_3$ and CrGeTe$_3$, respectively.

\begin{figure}[!hbp]
  \centering
  \includegraphics[scale=1.18,angle=0]{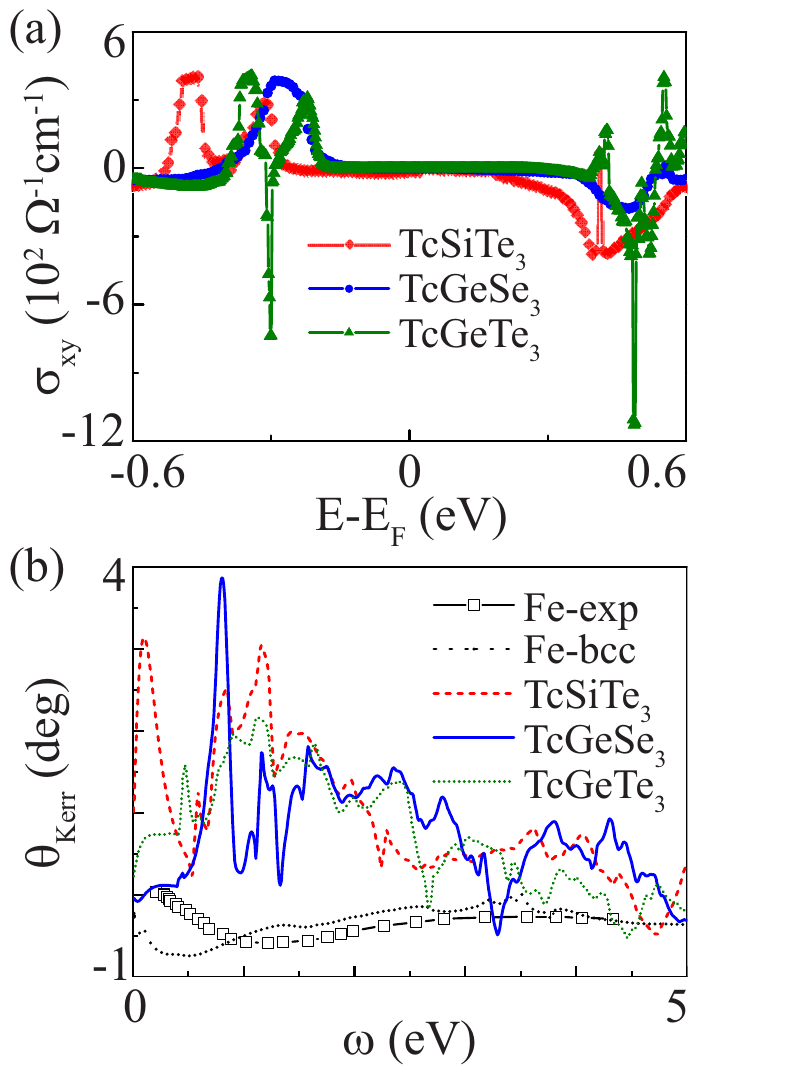}\\
  \caption{(a) The anomalous Hall conductivity of TcSiTe$_3$, TcGeSe$_3$ and TcGeTe$_3$ monolayers as a function of energy near Fermi level. (b) The Kerr angle $\theta_{Kerr}$ of TcSiTe$_3$, TcGeSe$_3$ and TcGeTe$_3$ monolayers as a function of photon energy, where the experimental (open squares) and calculated values (dotted line) of $\theta_{Kerr}$ for bulk Fe are included for comparison.}\label{fig5}
\end{figure}

Based on above Ising Hamiltonian and the estimated exchange parameter J, the Monte Carlo (MC) simulation was carried out to calculate the Curie temperatures of these 2D materials~\cite{Wolff1989}. The MC simulation was performed on a 60$\times$60 2D honeycomb lattice using 10$^6$ steps for each temperature. The magnetic moment as a function of temperature is shown in Fig.~\ref{fig4}(b). It can be seen that the normalized magnetic moment decreases rapidly to vanish at Curie temperature about 538 K, 212 K, 187 K and 19 K for TcSiTe$_3$, TcGeSe$_3$, TcGeTe$_3$ and CrGeTe$_3$ monolayers, respectively. The results indicate that TcSiTe$_3$, TcGeSe$_3$ and TcGeTe$_3$ monolayers can be potential candidates for high temperature 2D ferromagnetic semiconductors.

\begin{table*}[t]
	\caption{The dominant hopping matrix elements $\left|V\right|$ and energy difference $\left|E_p-E_d\right|$ between $p$ orbitals of Si (Te) and $d$ orbitals of Tc (Cr) in eV for TcSiTe$_3$ (CrGeTe$_3$) monolayer.}\label{tab:hopping}
	\begin{tabular}{c|c|c|c|c|c|c}
		\hline
        \hline
         monolayer     &  &$p_z$-$d_{z^2}$ &$p_z$-$d_{xz}$ &$p_z$-$d_{yz}$ &$p_z$-$d_{x^2-y^2}$ &$p_z$-$d_{xy}$\\
		\hline
		\multirow{2}{1cm}{TcSiTe$_3$}  &$\left|V\right|$    &0.444869     &0.096455     &0.254298 &0.179421    &0.386326     \\
        \cline{2-7}
                      &$\left|E_p-E_d\right|$ &0.158048     &0.118685     &0.445139     &0.296614 &0.071643\\
		\hline  \hline
		  &      &$p_x$-$d_{z^2}$ &$p_x$-$d_{xz}$ &$p_x$-$d_{yz}$ &$p_x$-$d_{x^2-y^2}$ &$p_x$-$d_{xy}$\\
		\hline
        \multirow{2}{1cm}{CrGeTe$_3$}   &$\left|V\right|$    &0.447729     &0.216707     &0.045947      &0.720887     &0.055202     \\
        \cline{2-7}
                     &$\left|E_p-E_d\right|$  &0.717856      &1.045652     &1.118030      &0.627825 &1.000211     \\
        \hline
	\end{tabular}
\end{table*}

Due to the large SOC in these ferromagnetic semiconductors, a large anomalous Hall conductivity (AHC) is expected. We calculated the intrinsic AHC due to the Berry curvature of electronic band structure as shown in Fig.~\ref{fig5}(a). The magnitude of AHC $\sigma_{xy}$ for the p-type TcGeTe$_3$ can reach 7.5$\times$10$^2$ ($\Omega$$\cdot$cm)$^{-1}$ and the n-type TcGeTe$_3$
can be up to 1.1$\times$10$^3$ ($\Omega$$\cdot$cm)$^{-1}$. The p- or n-type TcSiTe$_3$ and TcGeSe$_3$ can be as large as 4$\times$10$^2$ ($\Omega$$\cdot$cm)$^{-1}$. These values are comparable to the intrinsic $\sigma_{xy}$ in some ferromagnetic metals, such as 7.5$\times$10$^2$ ($\Omega$$\cdot$cm)$^{-1}$ in bcc Fe~\cite{Yao2004,Wang2006}, and 4.8$\times$10$^2$ ($\Omega$$\cdot$cm)$^{-1}$ in fcc Ni~\cite{Wang2007} due to the Berry curvature of band structures.

Large magneto-optical Kerr effect (MOKE) is also possible in 2D ferromagnetic materials with large SOC~\cite{Gu2017}. We investigated the MOKE for TcSiTe$_3$, TcGeSe$_3$ and TcGeTe$_3$ monolayers. The Kerr rotation angle is given by:
\begin{equation}\label{Kerr-angle}
\theta_{Kerr}(\omega) = -Re \frac{\epsilon_{xy}}{(\epsilon_{xx}-1)\sqrt{\epsilon_{xx}}},
\end{equation}
where $\epsilon_{xx}$ and $\epsilon_{xy}$ are the diagonal and off-diagonal components of the dielectric tensor $\epsilon$, and $\omega$ is the photon energy, respectively. The dielectric tensor $\epsilon$ can be obtained by the optical conductivity tensor,
$\sigma(\omega)$ = $\frac{\omega}{4\pi i}[\epsilon(\omega)-I]$, where $I$ is the unit tensor.
We performed the calculations with VASP along with the wannier90 tool to obtain the optical conductivity tensor $\sigma$ and the Kerr angle $\theta_{Kerr}$. The calculated $\theta_{Kerr}$ as a function of photon energy for TcSiTe$_3$, TcGeSe$_3$ and TcGeTe$_3$ monolayers is shown in Fig.~\ref{fig5}(b). Our calculated and previous experimental results for Fe metal are also included for comparison~\cite{krinchik1968}. It can be seen that a large Kerr angle $\theta_{Kerr}$ is obtained for TcSiTe$_3$, TcGeSe$_3$ and TcGeTe$_3$ monolayers, particularly for photon energies $\omega$ near 1 eV. The maximal Kerr angle for TcSiTe$_3$, TcGeSe$_3$ and TcGeTe$_3$ monolayers is an order of magnitude larger than that for CrGeTe$_3$ monolayer~\cite{Gong2017}, and about 5 times larger than that for bulk Fe.

\section{Discussion}
How to understand the enhanced Curie temperature of TcSiTe$_3$ monolayer compared with CrGeTe$_3$ monolayer? According to the superexchange interaction ~\cite{Goodenough1955,Kanamori1960,Anderson1959}, the FM coupling is expected since the Tc-Te-Tc and Cr-Te-Cr bond angles are close to 90 degree. The indirect FM coupling between Tc (Cr) atoms is proportional to the direct AFM coupling between neighboring Tc(Cr) and Te atoms. The magnitude of this direct AFM coupling can be roughly estimated as $J =\left|V\right|^2/\left|E_p-E_d\right|$, where $\left|V\right|$ is the hopping matrix element between $p$ orbitals of Te and $d$ orbitals of Tc (Cr), and $\left|E_p-E_d\right|$ is the energy difference between $p$ orbitals of Te and $d$ orbitals of Tc (Cr). By using maximally-localized Wannier orbital projections, the dominant hopping matrix elements $\left|V\right|$ and their corresponding energy differences $\left|E_p-E_d\right|$ can be obtained for 2D TcSiTe$_3$ and CrGeTe$_3$, respectively, as listed in Table~\ref{tab:hopping}. The results suggest that the direct AFM coupling for TcSiTe$_3$ is dominated by the $p_z$ orbitals of Te and $d_{z^2}$ and $d_{xy}$ orbitals of Tc. Because the $p_z$ orbitals of Te and $d_{z^2}$ and $d_{xy}$ orbitals of Tc for TcSiTe$_3$ monolayer are very close to each other in energy, and at the same time the sufficiently large hoppings exist between them, a large AFM coupling between Te and Tc atoms of TcSiTe$_3$ is obtained. Although the hopping parameter is quite large between the $p_x$ orbitals of Te and $d_{z^2}$ and  $d_{x^2-y^2}$ orbitals of Cr for CrGeTe$_3$ monolayer, because of the large energy differences among them, the AFM coupling between Te and Cr atoms is much weaker than that for TcSiTe$_3$ monolayer.

\begin{figure*}[tbhp]
  \centering
  \includegraphics[scale=0.5,angle=0]{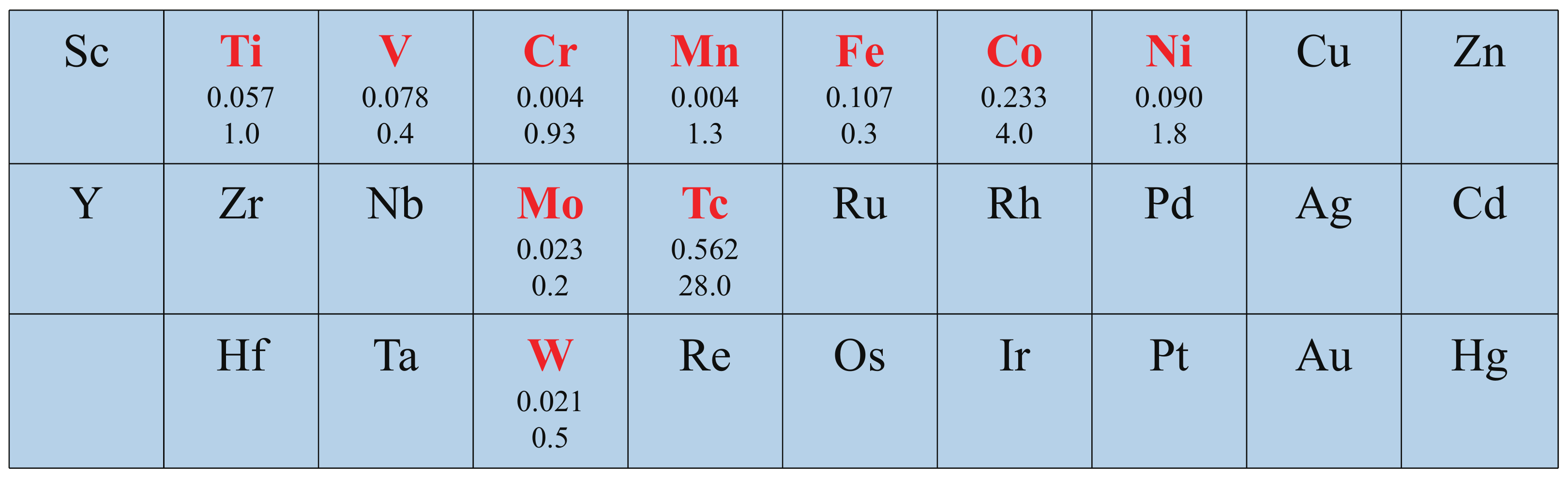}\\
  \caption{For MGeTe$_3$ (M=3d, 4d, 5d metals), the orbital moment (number in the first line, unit in $\mu_B$) and magnetocrystalline anisotropy energy (MAE, number in the second line, unit in meV). That the compound MGeTe$_3$ is a magnetic material is indicated with M in red color. The MAE is calculated by the energy difference per M atom between the FM$^z$ and FM$^x$ configurations.}\label{fig6}
\end{figure*}

Are the giant orbital moments in TcSiTe$_3$, TcGeSe$_3$ and TcGeTe$_3$ monolayers unique? To answer this question, we study the 2D MGeTe$_3$ monolayers with M = 3d, 4d, 5d transition metals. The results of orbital moment and magnetocrystalline anisotropy energy are listed in Fig.~\ref{fig6}.
By the spin-polarization calculations of these monolayers, only ten are found to be magnetic, and these magnetic materials are colored in red for M metals in Fig.~\ref{fig6}. Among these magnetic materials, the largest orbital moment is 0.53 $\mu_B$ in TcGeTe$_3$, two times larger than the 2nd largest orbital moment in CoGeTe$_3$, and about an order of magnitude larger than the orbital moment in the rest 2D materials. The same unique behavior of Tc is also found in the results of MAE. As listed in Fig.~\ref{fig6}, one may find that among these 2D MGeTe$_3$ materials, 2D TcGeTe$_3$ has an extraordinarily large MAE.

\section{Conclusion}
By first-principles calculations, we proposed three stable 2D Ising-type ferromagnetic semiconductors of TcSiTe$_3$, TcGeSe$_3$ and TcGeTe$_3$ with high Curie temperatures of 538 K, 212 K and 187 K, respectively. Due to large spin-orbit couplings, the large magnetocrystalline anisotropy energy, large anomalous Hall conductivity, and large magneto-optical Kerr effect are found in these intriguing 2D ferromagnetic semiconductors. Comparing all possible 2D MGeTe$_3$ materials (M = 3d, 4d, 5d transition metals), the unique behavior of Tc is highlighted with extraordinarily large orbital moment near 0.5 $\mu_B$. The large orbital moments are unveiled to be from the comparable crystal fields and electron correlations in these Tc-based 2D materials. The microscopic mechanism of the high Curie temperature is also addressed. Our findings present a new series of materials with large spin-orbit coupling that would have essential implications in designing spintronic devices for next generation microelectronics.

\section*{Acknowledgements}
B.G. is supported by the National Natural Science Foundation of China (Grant No. Y81Z01A1A9), the Chinese Academy of Sciences
(Grant No. Y929013EA2) and the University of Chinese Academy of Sciences (Grant No. 110200M208). G.S. is supported in part by the National Key R\&D Program of China (Grant No. 2018FYA0305800), the Strategic Priority Research Program of the Chinese Academy of Sciences
(Grants No. XDB28000000 and No. XBD07010100), the National Natural Science Foundation of China (Grant No. 11834014), and Beijing Municipal Science and Technology Commission (Grant No. Z118100004218001).

\ \
\par
\ \
\begin{appendix}
\renewcommand{\theequation}{A\arabic{equation}}
\setcounter{equation}{0}
\renewcommand{\thefigure}{A\arabic{figure}}
\setcounter{figure}{0}
\renewcommand{\thetable}{A\arabic{table}}
\setcounter{table}{0}


\end{appendix}

\newpage
\bibliographystyle{apsrev4-1}
%

\end{document}